# Numerical simulations of modulated waves in a higher-order Dysthe equation

**Alexey Slunyaev[1,4)] and Efim Pelinovsky[2-5)]**

1) Institute of Applied Physics, Nizhny Novgorod, Russia
2) Institute of Applied Physics, Nizhny Novgorod, Russia
3) Nizhny Novgorod State Technical University n.a. R. Alekseev, Nizhny Novgorod, Russia
4) National Research University – Higher School of Economics, Moscow, Russia
5) Special Research Bureau for Automatization of Marine Resechers, Yuzhno-Sakhalinsk, Russia

Alexey Slunyaev (Corresponding author):
ORCID: 0000-0001-7782-2991
Slunyaev@appl.sci-nnov.ru
Postal address: Institute of Applied Physics, Box-120, 46 Ulyanova Street, Nizhny Novgorod, 603950 Russia

Efim Pelinovsky:
ORCID: 0000-0002-5092-0302
Pelinovsky@appl.sci-nnov.ru
Postal address: Institute of Applied Physics, Box-120, 46 Ulyanova Street, Nizhny Novgorod, 603950 Russia

## Abstract

The nonlinear stage of the modulational (Benjamin – Feir) instability of unidirectional deep water surface gravity waves is simulated numerically by the firth-order nonlinear envelope equations. The conditions of steep and breaking waves are concerned. The results are compared with the solution of the full potential Euler equations and with the lower order envelope models (the 3-order nonlinear Schrödinger equation and the standard 4-order Dysthe equations). The generalized Dysthe model is shown to exhibit the tendency to re-stabilization of steep waves with respect to long perturbations.

## 1. Introduction

The concept of a wave envelope is very efficient in the situations of slow wave modulations, i.e., narrow spectra. It reduces the problem of description of individual waves to a simpler problem of description of the wave modulation. The nonlinear Schrödinger equation (NLS) for water waves was first derived by Benney & Newell (1967) and then by Zakharov (1968) and Hasimoto & Ono (1972) as the leading-order approximation for weakly nonlinear waves, with dominating four-wave interactions. The corresponding nonlinear term is of the order $O(\varepsilon^3)$, where the small parameter $\varepsilon \ll 1$ is related to the small wave steepness, $\varepsilon = kH/2$ ($k$ is the wavenumber and $H$ is the wave height). In this theory the dispersion is assumed to be of the same order of magnitude as the nonlinearity, $\Delta k/k = O(\varepsilon)$ ($\Delta k$ – is the characteristic width of the wavenumber spectrum). However, realistic sea waves are not narrow-banded, and the NLS equation turns out to be a rather rough model for the wave description; its practical applications are very limited due to this reason. Kristian Dysthe derived an improved theory for modulated waves in the deep water [Dysthe, 1979]. The new theory implied the same balance between the nonlinear and dispersive terms, though took into account the next order contributors of the nonlinear dispersion, $O(\varepsilon^4)$. It is more important that within the Dysthe model the wave components of the dominant wave length and the induced long-scale waves are treated separately.

The Dysthe equations were derived alternatively from the Zakharov equations in the narrow spectrum limit by Stiassnie (1984). Hence, the envelope theory possesses firm relations with the nonlinear spectral theory developed previously by Hasselmann and Zakharov [Hasselmann, 1962; Zakharov, 1968]. The dispersive part of the Dysthe equations was improved in Trulsen & Dysthe (1996) and Debsarma & Das (2005) assuming a different balance between the wave nonlinearity and dispersion ($\Delta k/k = O(\varepsilon^{1/2})$). For the purpose of numerical solution, the full linear dispersion may be easily taken into account as suggested in [Trulsen et al, 2000]. The fourth-order system of equations for wave modulations was derived for the finite-depth case by Brinch-Nielsen & Jonsson (1986) though the obtained theory was much more complicated than the one of Dysthe.

The Dysthe theory has received much recognition in the community of water wave modelers due to its efficiency in description of realistic sea waves (e.g. Shemer et al, 2002; Trulsen, 2007; Goullet & Choi, 2011), simplicity and high speed of numerical computations with the help of pseudo-spectral methods. Due to the asymptotic essence of the Dysthe theory, its modifications for the description of wave evolution in time and in space exist [Trulsen, 2006]. The latter is most convenient for the simulation of waves generated by wavemakers in laboratory tanks. The Dysthe equations have been used in a number of works of the authors of this paper, where they were compared with results of numerical simulations of the full equations and also with laboratory measurements (e.g., Slunyaev et al, 2013, 2014; Slunyaev, 2009; Shemer et al., 2010). In most of the cases the simulation of the Dysthe equations yielded from good to perfect results. The wave modulation recurrence phenomenon within the Dysthe equations was studied by Janssen (1983) and in the recent paper by Armaroli et al (2017).

A high order asymptotic NLS theory accurate to the order $O(\varepsilon^5)$ was derived in [Slunyaev, 2005] for unidirectional waves propagating in water with constant depth, assuming that the effects of nonlinearity and dispersion are of the same order of magnitude, $\Delta k/k = O(\varepsilon)$. The obtained nonlinear coefficients diverge in the limit of a large depth, hence this limit requires a more accurate approach using the idea of Dysthe to separate the part responsible for the induced flow. Then the Dysthe equations in the order $O(\varepsilon^5)$ were obtained in [Slunyaev, 2005] with some incomplete accounting for the mean flow effect (see the discussion in Sec. 2).

The focusing NLS equation and the Dysthe equations describe the modulational (Benjamin – Feir) instability of surface waves. The modulational instability and related breather solutions of the NLS equation were suggested to explain the rogue wave phenomenon (see e.g. Dysthe & Trulsen, 1999; Onorato et al, 2001, and the reviews Dysthe et al, 2008; Kharif et al, 2009; Slunyaev et al, 2011; Onorato et al, 2013; Dudley et al, 2014). The description of so-called abnormal or rogue waves is probably one of the most fascinating problems of the modern oceanography, and one of the most challenging applications of the nonlinear water wave theories. Rogue waves are frequently assumed to be caused by the nonlinear self-modulation processes which occur on the sea surface when favorable conditions are created. The strongly nonlinear stage of the modulational instability is the most interesting as then the wave dynamics may deviate essentially from the weakly nonlinear frameworks. The dynamics of unstable modulations and also breather solutions of the NLS equation have been tested against fully nonlinear simulations and laboratory measurements in plenty of publications [Henderson et al, 1999; Chabchoub et al, 2011, 2012; Shemer & Alperovich, 2013; Slunyaev & Shrira, 2013; Slunyaev et al, 2013; Slunyaev & Dosaev, 2019).

In this paper we present numerical simulations showing effects of the high order Dysthe equations derived in [Slunyaev, 2005]; modulationally unstable wave packets are simulated numerically. In Sec. 2 the classic and high order Dysthe models are briefly presented. The description of the numerical simulations of the nonlinear stage of the Benjamin – Feir instability is given in Sec. 3. Besides the extended Dysthe theory, the simulations are performed within different frameworks listed in this section. Brief conclusions are given in the end.

## 2. The classic and the next-order Dysthe equations

Let us assume that the waves propagate along the horizontal axis $Ox$, the axis $Oz$ is upward directed. The narrowness of the spectrum makes possible separation of different wave harmonics, having the carrier wavenumber $k_0$ and the carrier frequency $\omega_0$. The solution of the hydrodynamic equations with the help of asymptotic approach is based on a decomposition of the non-zero harmonics of the wave field, the surface displacement, $\eta(x, t)$, and the velocity potential, $\varphi(x, z, t)$, in the series of small parameter $\varepsilon$, which corresponds to the small nonlinearity and weak dispersion,

$$\eta = \overline{\eta}(x,t) + \sum_{m=1}^{\infty} \varepsilon^m \operatorname{Re} \sum_{n=1}^{\infty} A_{nm}(x,t) e^{in\omega_0 t - ink_0 x}, \tag{1}$$

$$\varphi = \overline{\varphi}(x,z,t) + \sum_{m=1}^{\infty} \varepsilon^m \operatorname{Re} \sum_{n=1}^{\infty} B_{nm}(x,t,z) e^{in\omega_0 t - ink_0 x}. \tag{2}$$

Note that in (1), (2) the real-valued mean flow $\overline{\varphi}$ and the long-scale elevation $\overline{\eta}$ are detached.

The assumptions of slow modulations and weak nonlinearity allow making distinction between fast and slow coordinates. Complex-valued amplitudes $A_{nm}$ and $B_{nm}$ are slow functions of the distance, $x$, and time. Due to the separation of scales, the Laplace equation applies to each of the harmonic of $\varphi$, and it possesses an analytic solution for the vertical structure of $B$ (see (14) below). The induced potential $\overline{\varphi}$ and the elevation $\overline{\eta}$ are slow functions of coordinates and time. The expansions (1) and (2) are substituted into the primitive potential hydrodynamic equations for planar waves which consist of two surface boundary conditions,

$$\frac{\partial \varphi}{\partial z} = \frac{\partial \eta}{\partial t} + \frac{\partial \varphi}{\partial x} \frac{\partial \eta}{\partial x}, \quad z = \eta(x,t), \tag{3}$$

$$\frac{\partial \varphi}{\partial t} + g\eta + \frac{1}{2}\left(\left(\frac{\partial \varphi}{\partial x}\right)^2 + \left(\frac{\partial \varphi}{\partial z}\right)^2\right) = 0, \quad z = \eta(x,t), \tag{4}$$

of the Laplace equation in the water bulk

$$\frac{\partial^2 \varphi}{\partial x^2} + \frac{\partial^2 \varphi}{\partial z^2} = 0, \quad -h \leq z \leq \eta, \tag{5}$$

and of the non-leaking bottom boundary condition

$$\frac{\partial \varphi}{\partial z} = 0, \quad z = -h. \tag{6}$$

Collecting the terms of different orders of $\varepsilon$, one obtains the relations between $A_{nm}$, $B_{nm}$, $\overline{\varphi}$, $\overline{\eta}$, and also the evolution equations (see e.g. [Trulsen, 2007]). When the terms smaller than $O(\varepsilon^3)$ are disregarded, one obtains nonlinear Schrödinger (NLS) equation

$$i\left(\frac{\partial A}{\partial t}+C_{gr}\frac{\partial A}{\partial x}\right)+\frac{\omega_0}{8k_0^2}\frac{\partial^2 A}{\partial x^2}+\frac{\omega_0 k_0^2}{2}|A|^2 A=0, \quad (7)$$

where $C_{gr} = \omega_0/k_0/2$ is the group velocity. In the next order $O(\varepsilon^4)$ the Dysthe system appears,

$$i\frac{\partial A}{\partial t}+\hat{F}^{-1}\{\hat{L}_k\hat{F}\{A\}\}+\frac{\omega_0 k_0^2}{2}|A|^2 A+i\frac{3\omega_0 k_0}{2}|A|^2\frac{\partial A}{\partial x}+i\frac{\omega_0 k_0}{4}A^2\frac{\partial A^*}{\partial x}+k_0 A\frac{\partial \bar{\phi}}{\partial x}=0; \quad (8)$$

$$\frac{\partial^2 \bar{\varphi}}{\partial x^2}+\frac{\partial^2 \bar{\varphi}}{\partial z^2}=0; \quad (9)$$

$$\frac{\partial \bar{\varphi}}{\partial z}=\frac{\omega_0}{2}\frac{\partial}{\partial x}|A|^2, \quad z=0; \quad (10)$$

$$\frac{\partial \bar{\varphi}}{\partial z}=0, \quad z=-h. \quad (11)$$

The first equation (8) is of the kind of the extended nonlinear Schrödinger equation for the complex amplitude $A(x,t)$ with a contribution of the induced flow potential calculated at the rest water level, $\bar{\phi}(x,t) \equiv \bar{\varphi}(x, z=0, t)$. $\hat{L}$ is the dispersion operator described below. The other part (9)-(11) represents the Laplace equation on the flow $\bar{\varphi}(x, z, t)$ in the water bulk with the upper surface nonlinear boundary condition defined at the rest water level, $z = 0$. The coefficients in (8)-(11) are calculated in the limit of very deep water; in this limit the bottom condition (6) may be replaced with $\bar{\varphi} \to 0$ as $z \to -\infty$.

The original Dysthe system [Dysthe, 1979] contains a few dispersive terms in (8), which take into account three orders of the Taylor expansion of the linear dispersive relation for waves in water of the depth $h$,

$$\omega(k)=\sqrt{gk\tanh(kh)}, \quad (12)$$

in the vicinity of the chosen carrier wave $\omega_0 = \omega(k_0)$. If the infinitely deep water condition is implied, $k_0 h \gg 1$, then $\omega_0 = (gk_0)^{1/2}$. The full linear dispersion (12) may be easily taken into account with the help of operator $\hat{L}$ which acts in the Fourier space [Trulsen et al, 2000],

$$\hat{L}_k=\sqrt{g(k-k_0)\tanh[(k-k_0)h]}, \quad (13)$$

where $k$ are the array of wavenumbers; $\hat{F}\{\cdot\}$ in (8) denotes the Fourier transform in space. The corresponding term in (8) formally contains terms beyond the approximation $O(\varepsilon^4)$. This modification does not lead to greater computational efforts when spectral or pseudo-spectral approaches are applied and used systematically below to solve each of the three envelope equations. If necessary, the full dispersion operator $\hat{L}$ (13) may be expanded into the Taylor series to the wanted order of accuracy (see (A.5) in the Appendix).

The physical fields, the surface displacement, $\eta(x, t)$, and the velocity potential, $\varphi(x, z, t)$, may be obtained with the help of the reconstruction formulas which take into account the induced component and three orders of the asymptotic expansions for the weak nonlinearity and slow modulations as follows,

$$\eta(x,t)=\bar{\eta}+\eta^{(1)}+\eta^{(2)}+\eta^{(3)}, \quad \varphi(x,z,t)=\bar{\varphi}+\varphi^{(1)}+\varphi^{(2)}+\varphi^{(3)}, \quad (14)$$

$$\bar{\eta}=\frac{1}{2\omega_0}\frac{\partial \bar{\phi}}{\partial x},$$

$$\eta^{(1)}=\mathrm{Re}(AE), \quad \eta^{(2)}=\frac{k_0}{2}\mathrm{Re}(A^2 E^2), \quad \eta^{(3)}=-\frac{1}{2}\mathrm{Im}\left(A\frac{\partial A}{\partial x}E^2\right)+\frac{3k_0^2}{8}\mathrm{Re}(A^3 E^3),$$

$$\varphi^{(1)}=-\frac{\omega_0}{k_0}\mathrm{Im}(AE)e^{k_0 z}, \quad \varphi^{(2)}=\frac{\omega_0}{2k_0^2}\mathrm{Re}\left[\frac{\partial A}{\partial x}E\right](1-2k_0 z)e^{k_0 z},$$

$$\varphi^{(3)} = \frac{\omega_0 k_0}{8} \operatorname{Im}[A|A|^2 E] e^{k_0 z} + \frac{\omega_0}{8 k_0^3} \operatorname{Im}\left[\frac{\partial^2 A}{\partial x^2} E\right](3 - 4 k_0 z + 4 k_0^2 z^2) e^{k_0 z},$$

$$E \equiv \exp(i\omega_0 t - i k_0 x).$$

In (14) $A(x, t) = A_{11} + A_{12} + A_{13}$ (see (1)). In the first approximation which corresponds to the NLS equation (7) the reconstruction formulas include formally only single terms, $\eta = \eta^{(1)}$ and $\varphi = \varphi^{(1)}$.

The next-order asymptotic expansions for weakly nonlinear slowly modulated waves were derived in [Slunyaev, 2005],

$$i\frac{\partial A}{\partial t} + \hat{F}^{-1}\{\hat{L}_k \hat{F}\{A\}\} + \frac{\omega_0 k_0^2}{2}|A|^2 A + i\frac{3\omega_0 k_0}{2}|A|^2 \frac{\partial A}{\partial x} + i\frac{\omega_0 k_0}{4} A^2 \frac{\partial A^*}{\partial x} +$$

$$+ \frac{\omega_0 k_0^4}{2}|A|^4 A - \frac{5\omega_0}{8}|A|^2 \frac{\partial^2 A}{\partial x^2} + \frac{3\omega_0}{32} A^2 \frac{\partial^2 A^*}{\partial x^2} - \frac{3\omega_0}{16} A \frac{\partial A}{\partial x}\frac{\partial A^*}{\partial x} - \frac{19}{32} A^*\left(\frac{\partial A}{\partial x}\right)^2 + \quad (15)$$

$$+ k_0 A \frac{\partial \overline{\phi}}{\partial x} = 0.$$

The first and last lines in (15) replicate the terms from the Dysthe equation (8). The new terms $O(\varepsilon^5)$ in (15) are of the higher order nonlinear dispersion and of the nonlinearity. The problem on the induced current, represented in (15) by the term $\overline{\phi} \equiv \overline{\varphi}(x, z = 0, t)$, is borrowed from the classic Dysthe model (9)-(11), what, strictly speaking, makes the equation (15) inconsistent from the point of view of the asymptotic theory. Accounting for the high-order corrections caused by the induced flow would make the equations significantly more complicated, and in particular less attractive to use. The consistent asymptotic theory was developed in [Slunyaev, 2005] for the finite depth case, when the induced velocity potential was decomposed in series assuming that the group lengths are larger than the water depth.

The solution for $\overline{\phi}$ (9)-(11) may be expressed in terms of the Hilbert transform, which in the Fourier domain yields the following relation for the mean current,

$$\frac{\partial \overline{\phi}}{\partial x} = i\frac{\omega_0}{2} \hat{F}^{-1}\{i|k|\hat{F}\{|\eta|^2\}\}. \qquad (16)$$

It corresponds to induced counter flows beneath nonlinear wave groups, and to wave group set-downs, according to the relation for $\overline{\eta}$ in (14).

The stability analysis of plane wave solutions of the listed above envelope models with respect to weak long perturbations is given in the Appendix, where the growth rate $\sigma(K, A_0)$ (A.17, A.16) is obtained for the high-order Dysthe equation (A.1-A.4), as well as the growth rates which follow from the classic Dysthe equation and from the NLS theory (A.17-A.19). Here $A_0$ denotes the amplitude of the plane wave, and $K$ is the wavenumber of the modulation. In Fig. 1a the growth rates versus dimensionless wavenumber of perturbation are shown for the dimensionless wave amplitude $k_0 A_0 = 0.1$. The dependencies of the growth rates on the wave steepness for a given perturbation length are shown in Fig. 1b. One may see from Fig. 1a that though the Dysthe theory leads to a smaller interval of unstable perturbation wavenumbers compared to the NLS equation, the high-order theory has a weak opposite effect, which grows when the wave amplitude further increases. Fig. 1b reveals a qualitative difference of the high-order theory in the range of large wave steepness compared to the NLS and the classic Dysthe models. While the growth rates of the NLS and Dysthe equations monotonically increase with the wave steepness, the high-order system exhibits the tendency to re-stabilization of too steep waves. In Fig. 1b we also plot by the dash-dotted line the growth rate for the high-order Dysthe model, when the term responsible for the induced flow is artificially put equal to zero, with the purpose to visualize the effect of this component. It follows that the induced flow mainly affects the amplitude threshold when the instability occurs; the effect of a wave re-stabilization at large wave steepness remains.

The high order Dysthe system (15-16) is rather straightforward to implement numerically similar to the classic Dysthe equations (8)-(11). However, it was never simulated before. A derivation of a high-order asymptotic theory technically represents a rather complicated task despite available automatic algebra systems. Besides, the methodology of the asymptotic technique admits variations which may sometimes lead to conflicting results (see the discussion in Slunyaev (2005)). Therefore besides the investigation of the role of higher order terms in the evolution equation, numerical tests may also be used to verify indirectly the correctness of coefficients in (15).

## 3. Numerical simulations of the nonlinear stage of the modulational instability

The process of wave self-modulation due to the modulational (Benjamin – Feir) instability is a particularly interesting example of nonlinear wave dynamics in deep water. In this section we simulate a few cases of modulationally unstable wave trains with the help of the high-order Dysthe theory (15-16) (hereafter referred to as HODysthe). The system of equations is solved with the help of the pseudo-spectral algorithm which uses the split-step-Fourier method for the integration in time (same as in [Lo & Mei, 1985]). The spatial periodic conditions are used. Due to the terms of higher order of nonlinearity in (15) compared to (8), one may anticipate that the code for the extended Dysthe theory is more prone to instabilities, hence the programming is more demanding. In the present simulations a partial de-aliasing procedure by padding and truncation [Canuto et al, 2010] is employed for each quadratic product.

The fully nonlinear simulation of the potential Euler equations in the conformal mapping representation of Dyachenko & Zakharov [Zakharov et al, 2002] (hereafter DZ) is used as the reference. Besides, the potential Euler equations are simulated using the strongly nonlinear high order spectral method which resolves up to 7-wave interactions [West et al, 1987] (hereafter HOSM). The results are also compared with the simulations of the standard classic Dysthe equations (8)-(11), and also of the integrable NLS framework (9).

The simulated wave conditions are realistic; they are close to the ones examined in [Dyachenko & Zakharov, 2005] within the fully nonlinear potential equations for water waves. The initial condition for the solvers of the potential Euler equations (the surface displacement, $\eta(x, t = 0)$, and the surface velocity potential, $\Phi(x, t = 0) = \varphi(x, z = \eta(x, t = 0), t = 0))$ is represented by a train of ten Stokes waves with a weak modulation (5%) of the wave amplitude. The Stokes wave solution is calculated numerically with high accuracy. The initial condition for the envelope equations $A(x, t = 0)$ is found for the given $\eta(x, 0)$ and $\Phi(x, 0)$ with the help of the iterative procedure which inverses the reconstruction formulas (14) (similar to the work by Trulsen (2001), see also Slunyaev et al (2014)). When the evolution of the envelope $A(x, t > 0)$ is computed, the surface displacement is reconstructed back according to the formulas (14). Note that the surface is calculated with the use of the same formulas (14) for all the envelope models (NLS, Dysthe, HODysthe), therefore the difference may be caused only by the different dynamics of the envelope $A(x, t)$ within these equations.

Two simulations when the initial perturbations grow due to the Benjamin – Feir instability and eventually much higher waves occur, which however do not reach the breaking onset are shown in Figs. 2, 3. The first case (Fig. 2) corresponds to the initial wave steepness $k_0H/2 = 0.07$, where $H$ is the wave height (i.e., the vertical distance between the wave crest and through). The maximum surface displacement in the simulated domain versus this value at $t = 0$ is given in Fig. 2a as the function of time for the simulations of the full potential Euler equations (DZ), and the three envelope models discussed above. All the models exhibit the self modulation and the associated significant wave amplification. In the simulation of the NLS equation the modulation grows faster and reaches somewhat smaller amplitude than in the other cases; a similar observation was reported in [Slunyaev & Shrira, 2013]. The Dysthe and HODysthe models reproduce the evolution of the wave amplification factor rather well, including the fast oscillations associated with the fast individual wave movement in a twice slower strongly modulated group (see the inset in Fig. 2a). The classic Dysthe model just slightly undervalues the wave maxima, while the high order Dysthe model overestimates them more noticeably.

The difference between the simulations is estimated with the root-mean-square difference between the water surface in the particular simulation, $\eta$, and the reference simulation of the DZ code, $\eta_{DZ}$, at a given time

$$\Delta(t) = \sqrt{\frac{\int (\eta - \eta_{DZ})^2 dx}{\int \eta_{DZ}^2 dx}} \cdot 100\% . \qquad (17)$$

The corresponding curves $\Delta(t)$ are shown in Fig. 2b. The accuracy of the NLS equation is obviously the worst. When the waves begin to grow considerably (after 200 wave periods $T_0 = 2\pi/\omega_0$, see Fig. 2a), the difference between the NLS equation and the full equations becomes unacceptable. The Dysthe model exhibits quite reasonable difference $\Delta$, which gradually increases during the wave modulation growth and then remains at the level about 10%. The error of the high order Dysthe model is initially smaller than of the standard Dysthe model, but grows when waves focus, and eventually noticeably exceeds the amount of the error of the classic Dysthe model. The surfaces at the moment of the maximum wave $\eta_{DZ}$ are shown in Fig. 2c. The snapshots shown in Fig. 2d correspond to the moment when the error of the HODysthe model reaches the local maximum (cf. Fig. 2a). The solution of the classic Dysthe equations in Fig. 2c,d seems to be closer to the simulation of the full equations (DZ), than the solution of the extended system. Meanwhile both the envelope models exhibit rather accurate description of waves.

A simulation of a steeper wave case $k_0 H/2 = 0.09$ is shown in Fig. 3. Similar to the previous case, before waves start to grow the error of the classic Dysthe model is larger than that of the high order Dysthe model (Fig. 3b). It increases quickly during the stage when large waves appear, and eventually reaches the value of the order of 40%. In the simulation of the high order Dysthe equation the error quickly grows after about 150 wave periods when the waves become steeper, and then the code blows up. Note that the simulation of the full equations does not reveal a wave breaking onset, hence the observed 'breaking' of the HODysthe simulation is not physical. The curve of the classic Dysthe model in Fig. 3a seems to agree with the fully nonlinear rather well, despite some noticeable underestimation of the wave amplitude during the extreme wave event. In the simulation of the NLS equation the waves grow faster and reach noticeably smaller amplitudes than in the other cases.

The snapshots of the surface displacement for two time instants are shown in Fig. 3c,d. The moment $t \approx 178 T_0$ shown in Fig. 3c is shortly before the simulation of the high order Dysthe equation stops, but the error of the simulation of the high order NLS equation $\Delta$ is already larger than the inaccuracy of the classic Dysthe model. The surfaces shown in Fig. 3d correspond to the moment of the maximum wave crest $\eta_{DZ}$ registered during the process of the wave self-modulation; it happens after the HODysthe code has stopped. It is obvious from Fig. 3c,d that the envelope models reproduce the wave evolution rather well, though the growing discrepancy of the HODysthe may be noticed in Fig. 3c, and some underestimation of the wave crest height by the Dysthe model is clearly observed in Fig. 3d. Note however that the vertical axis in Fig. 3c,d gives the measure of the wave steepness, which is quite high, bearing the weakly nonlinear essence of the models. As discussed in [Slunyaev & Shrira, 2014], the local steepness near the crest of the maximum amplified wave is even larger than that.

In the cases considered above (Figs. 2, 3) the simulations of the HOSM code do not show any noticeable disagreement with the simulations of the DZ code. The simulations of the NLS equation with the full linear dispersion (introduced with the help of the operator $\hat{L}$ (13)) are very close to the solutions of the integrable NLS equation (14).

The simulation of a different wave condition is shown in Fig. 4, where the initial wave steepness is further increased, $k_0H/2 = 0.15$ (thus the simulation conditions are the same as used in the study by Dyachenko & Zakharov (2005)). This case corresponds to the wave overturning, which cannot be simulated by the fully nonlinear code beyond some moment. As before, the NLS model represents very different results from the reference case of the fully nonlinear simulation (Fig. 4a). The simulation of the HOSM code blows up somewhat earlier than the breaking in the DZ code occurs (Fig. 4a,b); it generally agrees with the DZ code, though some negligible difference starts growing close to the breaking event (Fig. 4b). In this simulation of steeper waves the improvement of the high order Dysthe model with respect to the classic one is distinct in Fig. 4b. However close to the breaking event the error of the HODysthe code grows rapidly; the Dysthe and HODysthe codes exhibit similar errors approximately at the moment when the HOSM scheme breaks down (~$84T_0$). Before that the accuracy of the high order Dysthe model is much better than of the lower order equations (see also the inset in Fig. 4a). The instants before the breaking are shown in Fig. 4c,d, where only the surfaces in the vicinity of the maximum waves are shown. Some ripples in the surface simulated within the HODysthe code may be seen in Fig. 4c, which manifest the onset of numerical instability. At this moment the difference $\Delta$ for the high order Dysthe model is still much smaller than the inaccuracy of the classical Dysthe model, what may be concluded from Fig. 4c by a naked eye. The HODysthe model better reproduces the tendency of the maximum wave to skew in the onward direction (to the right). For the instant shown in Fig. 4d the solution of the HODysthe model is completely erroneous, while the solver of the Dysthe equations remains stable. The surfaces simulated by the HOSM and DZ codes well agree in this figure, though some noisy trembling of the curve for the HOSM may be seen. The standard Dysthe model turns out to be capable of reasonably accurate description of this near-breaking wave, though it does not reproduce the asymmetry of the maximum wave which tends to break.

## 4. Conclusion

The problem of definitively insufficient accuracy of the nonlinear Schrodinger equation for the description of nonlinear water waves was emphasized in the original paper by Kristian Dysthe [Dysthe, 1979]. Remarkably different growth rates of the Benjamin – Feir instability for waves with the steepness $\varepsilon = kH/2 > 0.1$ in the exact equations and in the NLS framework were particularly noted. The extension of the NLS theory suggested in [Dysthe, 1979] helped to improve the accuracy of the envelope model drastically. Today this system of equations is called the Dysthe equations; it has received some further development.

In this paper the nonlinear stage of the modulational (Benjamin – Feir) instability is simulated numerically in several asymptotic envelope equations of different orders of accuracy, the strongly nonlinear high order spectral method and full potential Euler equations. The focus is made on solving the high order Dysthe equation derived in [Slunyaev, 2005]. The simulated wave conditions correspond to unidirectional waves in deep water with initially moderate steepness ($k_0H/2 = 0.07...0.15$), while in the course of the unstable wave growth from steep to breaking waves appear. The simulation of the nonlinear Schrödinger equation yields too different results from the reference case of the simulation of the full potential Euler equations in conformal variables, hence the simulations of the more advanced models, the classic Dysthe equations and the high order model are considered more intently. Both the envelope models take into account the full linear dispersion law.

In all the simulated cases the classic fourth-order Dysthe equations seem to be more advantageous than the high order model. The HODysthe code exhibits smaller deviation from the fully nonlinear framework at the initial (though long) stage of the wave evolution. The difference in the inaccuracies is most evident when the waves are steep and tend to break. However when large strongly modulated waves appear due to the self-modulation effect, the inaccuracy of the HODysthe model rapidly increases, and the simulation blows up even if the waves do not break physically. At the present stage we cannot state whether the blow up results from the insufficiently accurate numerical simulation or it is an intrinsic property of the equation. The code for the high order Dysthe equations is noticeably more involved due to the higher order of nonlinearity and thus demonstrates worse stability properties. The computation time of the HODysthe code is inadequately large compared to the other simulated algorithms.

At the same time the better accuracy of the high order Dysthe equations at relatively small times may be treated as an implicit confirmation of consistency of the model derived in [Slunyaev, 2005]. At the early stage of the wave evolution the modulation is very slow and the induced mean flow is weak, see (10). When the growing group becomes shorter, the effect of the induced flow increases and hence its oversimplified description in the adopted high order model may become more dramatic. In the first place, this drawback should affect the wave group set-down and the group velocity. In Fig. 5 the profiles of the wave envelopes $|A(x)|$ calculated in the standard and extended Dysthe models are shown with solid lines; they correspond to the surfaces shown in Fig. 4c at the moment of time shortly before the breaking of the HODysthe code. It is clear that the envelope exhibits stronger front-tail asymmetry when simulated in the high order model (the capability of the Dysthe equation to describe the wave group asymmetry is well known, see e.g. Shemer et al (2002)). The domain shown in Fig. 5 initially contains 10 waves, hence the length of the focused group is about 2-3 dominant wavelength. The complex phase of the envelope $A(x)$ essentially influences the shape of individual waves. In Fig. 5 the dashed lines represent the real parts of $A(x)$ for the two simulated envelope models. It is obvious that besides the front-tail asymmetry of the envelopes, the complex phases are strongly asymmetric as well, and this asymmetry is even more pronounced in the high order framework. As a result, the high order model demonstrates a stronger front-back asymmetry of the individual maximum wave in the group, which tends to break (in fact, even stronger than in the full equations, see Fig. 4c). This property of the high order Dysthe model may be useful for better understanding of the processes associated with the wave skewing before it breaks.

A new direction in the development of envelope approaches for water waves may be noted, which is based on the Hamiltonian formalism and the Zakharov equation [Dyachenko et al, 2017a,b]. There new canonical variables and reduced kernels of nonlinear wave interactions are used to obtain so-called compact equations which govern the complex envelope functions with no assumption of narrow bandwidth. So far these models are limited to four-wave interactions, hence remain weakly nonlinear.

**Acknowledgements**

The support from the Russian Foundation for Basic Research (grant No. 16-55-52019) and from the Fundamental Research Programme of RAS "Nonlinear Dynamics" is acknowledged. EP conducted the research within the State Programme for Science (Task

№ 5.5176.2017/8.9), and was supported by the President Grant for the Leading Scientific Schools of the Russian Federation (NSh-2685.2018.5).

**Appendix. Modulational instability analysis**

In this section we transform the high-order evolution equations (9-11) and (15) to the form

$$i\left(\frac{\partial A}{\partial t}+C_{gr}\frac{\partial A}{\partial x}\right)+\beta_1\frac{\partial^2 A}{\partial x^2}+\alpha_1|A|^2 A+i\beta_2\frac{\partial^3 A}{\partial x^3}+i\alpha_{21}|A|^2\frac{\partial A}{\partial x}+i\alpha_{22}A^2\frac{\partial A^*}{\partial x}+ \quad (A.1)$$

$$+\beta_3\frac{\partial^4 A}{\partial x^4}+\alpha_{31}|A|^4 A+\alpha_{32}|A|^2\frac{\partial^2 A}{\partial x^2}+\alpha_{33}A^2\frac{\partial^2 A^*}{\partial x^2}+\alpha_{34}A\frac{\partial A}{\partial x}\frac{\partial A^*}{\partial x}+\alpha_{35}A^*\left(\frac{\partial A}{\partial x}\right)^2+k_0 A\frac{\partial\overline{\phi}}{\partial x}=0;$$

$$\frac{\partial^2\overline{\varphi}}{\partial x^2}+\frac{\partial^2\overline{\varphi}}{\partial z^2}=0; \quad (A.2)$$

$$\frac{\partial\overline{\varphi}}{\partial z}=\frac{\omega_0}{2}\frac{\partial}{\partial x}|A|^2, \quad z=0; \quad (A.3)$$

$$\frac{\partial\overline{\varphi}}{\partial z}\to 0, \quad z\to-\infty. \quad (A.4)$$

Here we introduce real-valued coefficients $\alpha_1$, $\alpha_{21}$, $\alpha_{22}$, $\alpha_{31}$, $\alpha_{32}$, $\alpha_{33}$, $\alpha_{34}$, $\alpha_{35}$ which are defined according to (15). The operator of the full linear dispersion is replaced with its expansion in the limit of infinite depth,

$$\hat{F}^{-1}\{\hat{L}_k\hat{F}\{A\}\}=iC_{gr}\frac{\partial A}{\partial x}+\beta_1\frac{\partial^2 A}{\partial x^2}+i\beta_2\frac{\partial^3 A}{\partial x^3}+\beta_3\frac{\partial^4 A}{\partial x^4}, \quad (A.5)$$

$$C_{gr}=\frac{\omega_0}{2k_0}, \quad \beta_1=\frac{\omega_0}{8k_0^2}, \quad \beta_2=-\frac{\omega_0}{16k_0^3}, \quad \beta_3=-\frac{5\omega_0}{128k_0^4}. \quad (A6)$$

The analysis of stability of the plane wave solution with respect to weak long modulations is performed following the usual way. The perturbed plane wave solution is sought in the form

$$A=A_0(1+\mu a(x,t))\exp(i\omega t+i\mu\psi(x,t)), \quad (A.7)$$

where $a(x,t)$ and $\psi(x,t)$ are unknown real functions with a small factor $\mu\ll 1$, $A_0$ is real and $\omega$ is the frequency of the unperturbed solution. The unknown variables will be sought in the form of harmonic functions with the perturbation wavenumber $K$ and the perturbation frequency $\Omega$ which may be complex-valued,

$$a=a_0\exp(i\Omega t-iKx)+c.c., \quad \psi=\psi_0\exp(i\Omega t-iKx)+c.c. \quad (A.8)$$

We solve the Laplace equation first. For the plane wave the boundary condition (A.3) results in the zero solution for the mean current. The solution for the perturbed wave will be sought in the form

$$\overline{\varphi}(x,z,t)=\mu\overline{\phi}(x,t)e^{|K|z}, \quad \overline{\phi}(x,t)=B\exp(i\Omega t-iKx)+c.c., \quad (A.9)$$

where $B$ is a constant. The ansatz (A.9) automatically satisfies (A.2) and (A.4). Inserting (A.9) and (A.7) into (A.3), in the order $O(\mu)$ we have

$$|K|(B\exp(i\Omega t-iKx)+c.c.)=\omega_0 A_0^2\frac{\partial a}{\partial x}. \quad (A.10)$$

Using the expression (A.8), we relate the amplitude of the envelope modulation and the large-scale velocity potential,

$$B=-i\,\text{sgn}(K)\omega_0 A_0^2 a_0. \quad (A.11)$$

Now we plug (A.8) into (A.1) and collect the terms at different powers of $\mu$. In the order $O(\mu)$ the frequency of the plane wave solution is determined,

$$\omega=\alpha_1 A_0^2+\alpha_{31}A_0^4. \quad (A.12)$$

In the next order $O(\mu^2)$ we obtain the relation on $a$ and $\psi$. Their real and imaginary parts give two relations,

$$a_t + C_{gr}a_x + \beta_1\psi_{xx} + \beta_2 a_{xxx} + \beta_3\psi_{xxxx} + (\alpha_{21} + \alpha_{22})A_0^2 a_x + \alpha_{32}A_0^2\psi_{xx} - \alpha_{33}A_0^2\psi_{xx} = 0, \quad (A.13)$$

$$-\psi_t - C_{gr}\psi_x + \beta_1 a_{xx} - \beta_2\psi_{xxx} + \beta_3 a_{xxxx} + 2\alpha_1 A_0^2 a + (-\alpha_{21} + \alpha_{22})A_0^2\psi_x +$$
$$+ 4\alpha_{31}A_0^4 a + \alpha_{32}A_0^2 a_{xx} + \alpha_{33}A_0^2 a_{xx} + k_0\overline{\phi}_x = 0. \quad (A.14)$$

With the use of (A.8), (A.9) and (A.11), equations (A.13) and (A.14) yield the compatibility condition

$$\Omega = C_{gr}K - \beta_2 K^3 + \alpha_{21}A_0^2 K \pm K\sqrt{D}, \quad (A.15)$$

$$D = \beta_1^2 K^2 - 2\beta_1\beta_3 K^4 + \beta_3^2 K^6 +$$
$$+ \left(-2\alpha_1\beta_1 + 2(\alpha_1\beta_3 + \alpha_{32}\beta_1)K^2 - 2\alpha_{32}\beta_3 K^4 + k_0\omega_0|K|(\beta_1 - \beta_3 K^2)\right)A_0^2 +$$
$$+ \left(2\alpha_1(-\alpha_{32} + \alpha_{33}) - 4\alpha_{31}\beta_1 + \alpha_{22}^2 + (4\alpha_{31}\beta_3 + \alpha_{32}^2 - \alpha_{33}^2)K^2 + k_0\omega_0|K|(\alpha_{32} - \alpha_{33})\right)A_0^4 +$$
$$+ 4\alpha_{31}(-\alpha_{32} + \alpha_{33})A_0^6. \quad (A.16)$$

The modulational instability occurs when $D < 0$, with the growth rate
$$\sigma = K\sqrt{-D}. \quad (A.17)$$

In the classic NLS theory (Eq. (7)) only the coefficients $C_{gr}$, $\alpha_1$ and $\beta_1$ are taken into consideration, and then

$$D_{NLS} = \beta_1(\beta_1 K^2 - 2\alpha_1 A_0^2), \quad (A.18)$$

what is a well-known result. Waves may be modulationally unstable for sufficiently long perturbations or large enough waves, if $\alpha_1\beta_1 > 0$, what is the case of deep water gravity waves.

For the standard Dysthe equation one should put $\alpha_{31} = 0$, $\alpha_{32} = 0$, $\alpha_{33} = 0$, $\alpha_{34} = 0$, $\alpha_{35} = 0$, $\beta_3 = 0$, and then (A.16) gives

$$D_{Dysthe} = \beta_1\left(\beta_1 K^2 - (2\alpha_1 - k_0\omega_0|K|)A_0^2\right) + \alpha_{22}^2 A_0^4. \quad (A.19)$$

This expression agrees with the one obtained in [Trulsen et al, 2000].

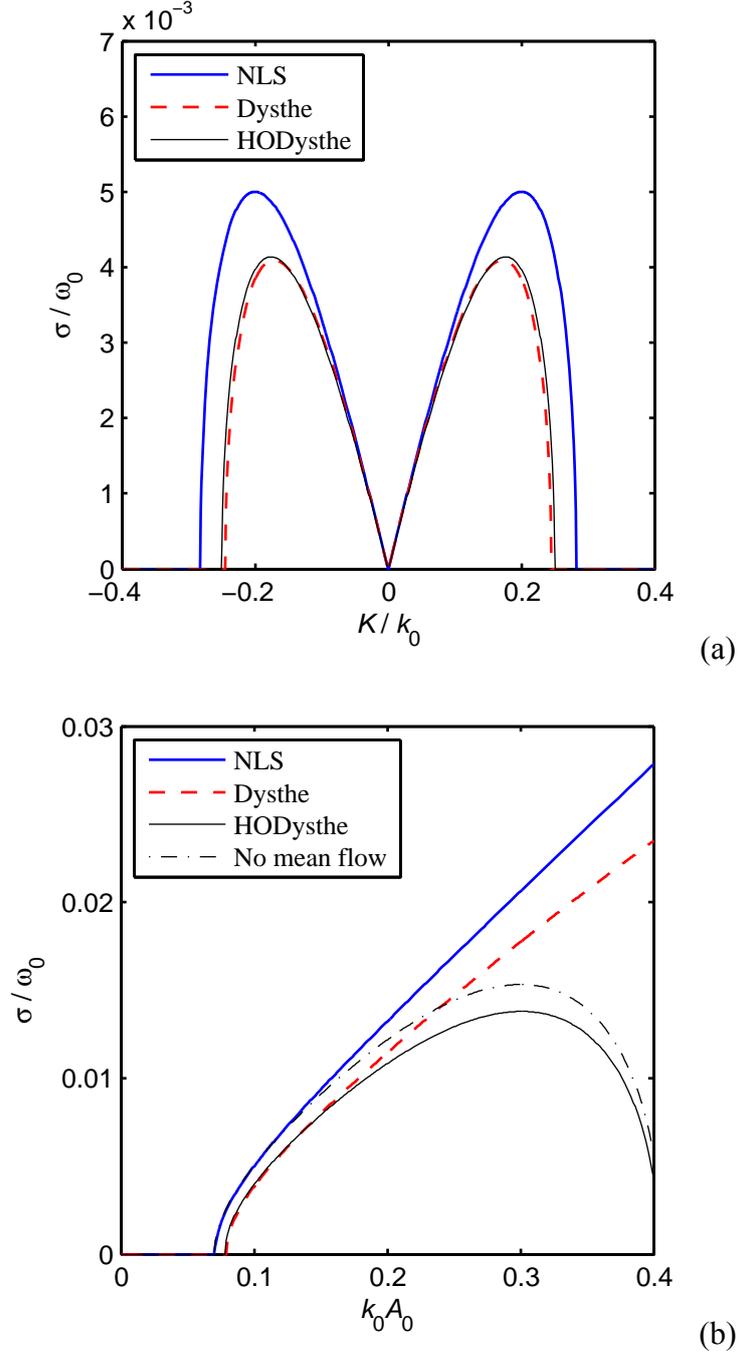

**Fig. 1.** Growth rates of the modulational instability in different envelope models as the functions of the perturbation wavenumber $K$ for the steepness $k_0A_0 = 0.1$ (a); and as the functions of the wave steepness for the perturbation $K = 0.2k_0$ (b).

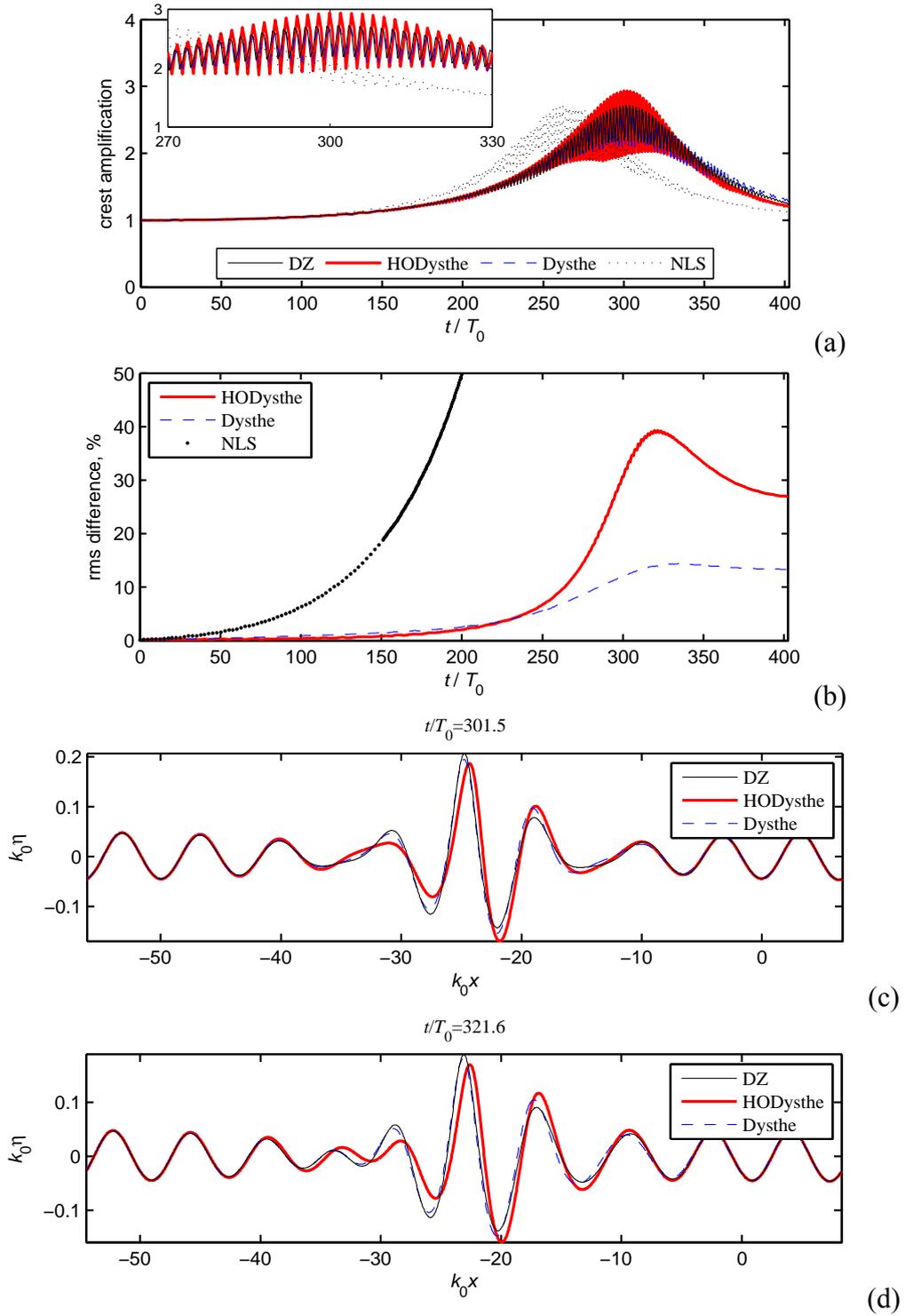

**Fig. 2.** Numerical simulation of a modulated train of 10 Stokes waves with the steepness $k_0H/2 = 0.07$ within different frameworks: the evolution of the maximum wave crest amplification (a); the evolution of the root-mean-square difference with respect to the reference simulation DZ (b); and the snapshots of the surface displacements at the moment of the maximum wave crest amplification (c) and at the moment of the maximum error of the HODysthe method (d).

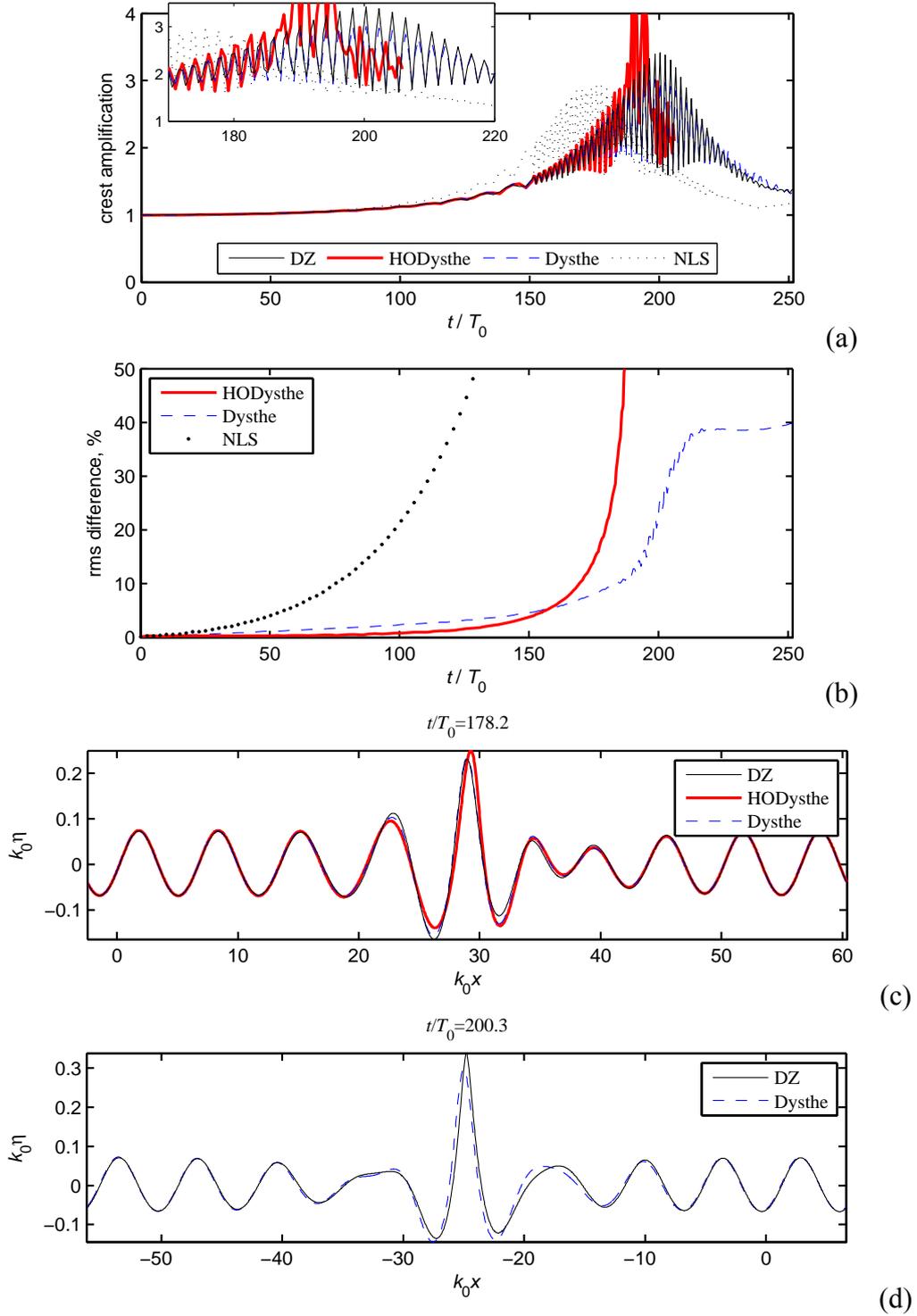

**Fig. 3.** Numerical simulation of a modulated train of 10 Stokes waves with the steepness $k_0H/2 = 0.09$ within different frameworks: the evolution of the maximum wave crest amplification (a); the evolution of the root-mean-square difference with respect to the reference simulation DZ (b); and the snapshots of the surface displacements at two instants (c, d).

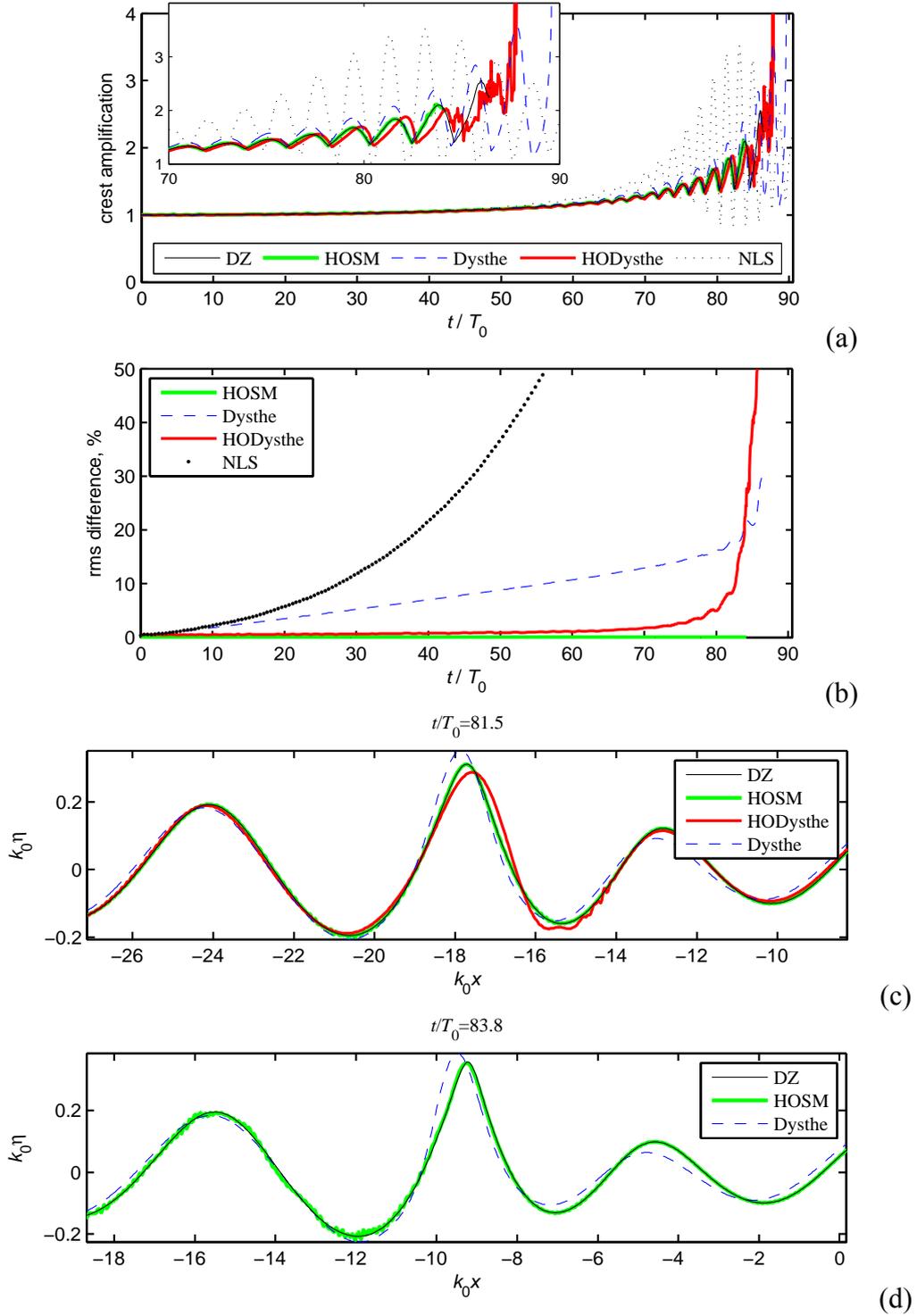

**Fig. 4.** Numerical simulation of a modulated train of 10 Stokes waves with the steepness $k_0H/2 = 0.15$ within different frameworks: the evolution of the maximum wave crest amplification (a); the evolution of the root-mean-square difference with respect to the reference simulation DZ (b); and the snapshots of the surface displacements at two instants in the vicinity of the maximum waves (c, d).

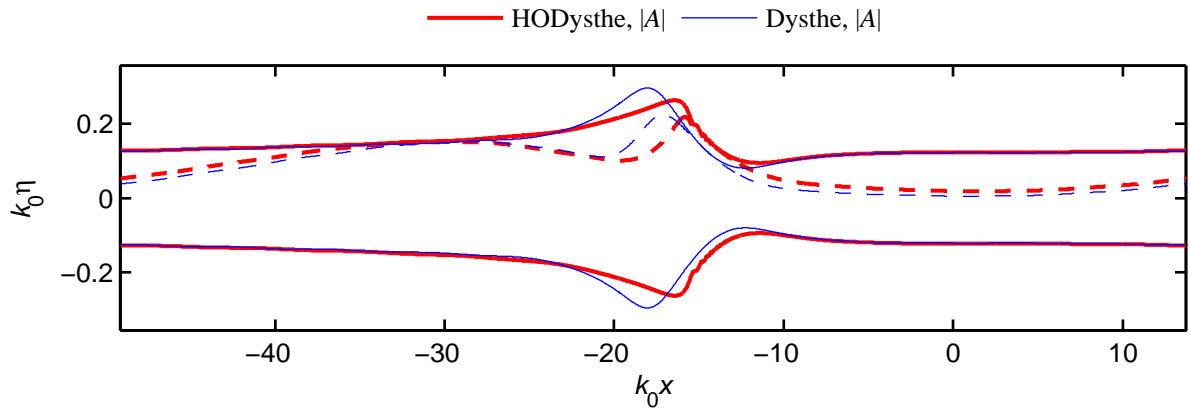

**Fig. 5.** The wave envelopes, |*A*| (solid curves), and the real parts of the complex amplitudes Re(*A*) (dashed lines), simulated within the Dysthe and the high order Dysthe modes shown for the instant $t = 81.5T_0$.